\documentclass[english,showpacs]{revtex4}
\usepackage[T1]{fontenc}
\usepackage[latin9]{inputenc}
\usepackage{textcomp}
\usepackage{amsmath}
\usepackage{amssymb}
\usepackage{esint}

\makeatletter
\@ifundefined{textcolor}{}
{%
 \definecolor{BLACK}{gray}{0}
 \definecolor{WHITE}{gray}{1}
 \definecolor{RED}{rgb}{1,0,0}
 \definecolor{GREEN}{rgb}{0,1,0}
 \definecolor{BLUE}{rgb}{0,0,1}
 \definecolor{CYAN}{cmyk}{1,0,0,0}
 \definecolor{MAGENTA}{cmyk}{0,1,0,0}
 \definecolor{YELLOW}{cmyk}{0,0,1,0}
}

\usepackage{latexsym}

\makeatother

\usepackage{babel}
\begin{document}

\title{On the Komar Energy and the Generalized Smarr Formula for a Charged Black Hole
of Noncommutative Geometry}

\author{Alexis Larrañaga}

\address{National Astronomical Observatory. National University of Colombia.
Bogota, Colombia.}

\email{ealarranaga@unal.edu.co}

\selectlanguage{english}%

\author{Juan Carlos Jimenez}

\address{Department of Physics. National University of Colombia.Bogota, Colombia.}
\begin{abstract}
\noindent We calculate the Komar energy $E$ for a 
charged black hole inspired by noncommutative geometry and identify the total mass ($M_{0}$) by considering
the asymptotic limit. We also found the generalized Smarr formula, which shows a deformation from the well known relation $M_{0}-\frac{Q_{0}^{2}}{r}=2ST$ depending on the noncommutative
scale length $\ell$ .
\end{abstract}

\maketitle
keywords: quantum aspects of black holes, thermodynamics, noncommutative geometry 

\section{Introduction}

There is a deep connection between gravity and thermodynamics that
has been  known for a long time, from the works of Bekenstein and
Hawking \cite{Beken3,Hawking1,Hawking2} to the recent research of
Padmanabhan \cite{padma,paddy}. In a thermodynamical system like
Schwarzschild black hole, the entropy $S$, Hawking temperature $T$
and energy $E$ are related by the first law of thermodynamics 
\begin{eqnarray}
dE & = & TdS,\label{intro1}
\end{eqnarray}

\noindent where $E$ is identified with the Komar energy \cite{komar,wald}
and specifically for a Schwarzschild black hole it equals the total
mass of the black hole, $M$. There is also an integral version of
this equation 
\begin{eqnarray}
E & =M= & 2TS.\label{intro2}
\end{eqnarray}
 known as the Smarr formula \cite{smarr} and it can be verified by
putting the expressions for entropy and the temperature 

\noindent 
\begin{equation}
T=\frac{1}{8\pi M}
\end{equation}

\noindent 
\begin{equation}
S=\frac{A}{4}=4\pi M^{2}.
\end{equation}

Eq. (\ref{intro2}) has been obtained in different ways \cite{padma,rbstat}
and the Komar energy is identified with the conserved charge associated
with the Killing vector defined at the event horizon (see for example
\cite{bmms}). Recently, some generalised expressions for Smarr formula
in different spacetimes have been studied \cite{rbstat,bmms,ms} and
in particular, the Kerr-Newman black hole with electric charge $Q$
and angular momentum $J$ satisfies the Smarr relation\cite{Poisson}

\begin{equation}
M=2TS+\Phi_{H}Q+2\Omega_{H}J\label{eq:generalSmarr}
\end{equation}
where $\Phi_{H}$ and $\Omega_{H}$ are the electric potential and
angular velocity at the horizon, respectively.

In this paper we investigate the specific case of a spherically symmetric
charged black hole inspired by noncommutative geometry \cite{NC1,NC2,NC3,NC4,NC5,spal,spal2,spalreview}.
This solution is obtained by introducing the noncommutativity effect
through a coherent state formalism \cite{sunfgs,sunPLB,rbreview},
which implies the replacement of the point distributions by smeared
structures throughout a region of linear size $\ell$. We perform
the analysis by obtaining the Komar energy by direct integration and
found the generalized Smarr formula, which shows a deformation from
the usual relation depending on the noncommutative parameter $\ell$.

\section{Komar Energy of the Charged Noncommutative Black Hole}

Many formulations of noncommutative fi{}eld theory are based on the
Weyl-Wigner-Moyal \textasteriskcentered{}-product \cite{moyal1,moyal2,moyal3}
that lead to some important problems such as Lorentz invariance breaking,
loss of unitarity or UV divergences of the quantum fi{}eld theory.
However, Smailagic and Spallucci \cite{NC1,NC2,NC3,NC4,NC5,spalreview}
explained recently a model of noncommutativity that can be free from
the problems mentioned above. They assume that a point-like mass $M$
and charge $Q$, instead of being quite localized at a point, must
be described by a smeared structure throughout a region of linear
size $\ell$. The metric for this distribution is given by, \cite{spal2},

\begin{equation}
ds^{2}=-f\left(r\right)dt^{2}+\frac{dr^{2}}{f\left(r\right)}+r^{2}d\Omega^{2}\label{eq:metric}
\end{equation}
where

\begin{equation}
f(r)=1-\frac{2M\left(r\right)}{r}+\frac{Q^{2}\left(r\right)}{r^{2}}\label{metric_coef}
\end{equation}

\begin{eqnarray}
Q\left(r\right) & = & \frac{Q_{0}}{\sqrt{\pi}}\sqrt{\gamma^{2}\left(\frac{1}{2},\frac{r^{2}}{4\ell^{2}}\right)-\frac{r}{\sqrt{2}\ell}\gamma\left(\frac{1}{2},\frac{r^{2}}{2\ell^{2}}\right)+\frac{\sqrt{2}r}{\ell}\gamma\left(\frac{3}{2},\frac{r^{2}}{4\ell^{2}}\right)}\label{eq:Q}\\
M\left(r\right) & = & \frac{2M_{0}}{\sqrt{\pi}}\gamma\left(\frac{3}{2},\frac{r^{2}}{4\ell^{2}}\right)\label{eq:M}
\end{eqnarray}
 and 
\begin{eqnarray}
\gamma\left(\frac{a}{b},x\right) & = & \int_{0}^{x}duu^{\frac{a}{b}-1}e^{-u}\label{incom_gamma}
\end{eqnarray}
 is the lower incomplete gamma function. Considering a spatial 2-sphere
$V$ with boundary $\partial V$, the Komar integral for the energy
is

\begin{equation}
E\left(V\right)=\frac{a}{16\pi}\oint_{\partial V}\nabla^{\mu}\xi^{\nu}d\Sigma_{\mu\nu}
\end{equation}
where the killing vector is $\xi=\frac{\partial}{\partial t}$, $d\Sigma_{\mu\nu}$
is the surface element at the boundary and the value of constant $a$
will be found by comparison with the noncommutative Schwarzschild
case. This is 

\begin{equation}
E=\frac{2a}{16\pi}\oint_{\partial V}\nabla^{\mu}\xi^{t}d\Sigma_{\mu t},
\end{equation}
where the factor $2$ appears because of the symmetry of the integrand.
The covariant derivative involved is
\begin{eqnarray}
\nabla_{\mu}\xi^{t} & = & \partial_{\mu}\xi^{t}+\Gamma_{\mu\sigma}^{t}\xi^{\sigma}=\Gamma_{\mu t}^{t},\label{eq:aux57}
\end{eqnarray}
and for the noncommutative charged solution the nonvanishing connections
are

\begin{eqnarray}
\Gamma_{rt}^{t} & = & \frac{-\frac{dM}{dr}r^{2}+rM+\frac{r}{2}\frac{dQ^{2}}{dr}-Q^{2}}{r\left(r^{2}-2Mr+Q^{2}\right)}\\
\Gamma_{tt}^{t} & = & \Gamma_{\theta t}^{t}=\Gamma_{\varphi t}^{t}=0,
\end{eqnarray}
 giving

\begin{eqnarray}
E & = & \frac{a}{8\pi}\oint_{\partial V}\frac{-\frac{dM}{dr}r^{2}+rM+\frac{r}{2}\frac{dQ^{2}}{dr}-Q^{2}}{r^{3}}d\Sigma_{rt}.
\end{eqnarray}

The surface element corresponds to

\begin{equation}
d\Sigma_{rt}=-d\Sigma_{tr}=-r^{2}\sin^{2}\theta d\theta d\varphi.
\end{equation}
and therefore

\begin{equation}
E=-\frac{a}{8\pi}\frac{-\frac{dM}{dr}r^{2}+rM+\frac{r}{2}\frac{dQ^{2}}{dr}-Q^{2}}{r}\oint_{\partial V}\sin^{2}\theta d\theta d\varphi
\end{equation}

\begin{equation}
E=\frac{a}{2}\left[\frac{dM}{dr}r-M-\frac{1}{2}\frac{dQ^{2}}{dr}+\frac{Q^{2}}{r}\right].
\end{equation}

By comparison with the Komar energy of the Schwarzschild black hole,
we shall identify $a=-2$. Hence, the energy of the noncommutative
charged black hole is finally given by 

\begin{equation}
E=M-\frac{dM}{dr}r-\frac{Q^{2}}{r}+Q\frac{dQ}{dr}.\label{eq:KomarEnergy}
\end{equation}

The horizons of the metric (\ref{eq:metric}) can be found by setting
$f(r_{\pm})=0$ , i.e.

\noindent 
\begin{equation}
r_{\pm}^{2}-2r_{\pm}M\left(r_{\pm}\right)+Q^{2}\left(r_{\pm}\right)=0,\label{eq:def_hor}
\end{equation}

\noindent which can be written as
\begin{equation}
r_{\pm}=M\left(r_{\pm}\right)\pm\sqrt{M^{2}\left(r_{\pm}\right)-Q^{2}\left(r_{\pm}\right)}.\label{eq:horizonsDef}
\end{equation}

Hawking temperature is defined in terms of the surface gravity at
the event horizon by

\begin{eqnarray}
T=\frac{\kappa}{2\pi} & = & \frac{1}{4\pi}\left.\partial_{r}f\left(r\right)\right|_{r=r_{+}},\label{Hawk_temp}
\end{eqnarray}
which gives in this case

\noindent 
\begin{eqnarray}
T & = & \frac{1}{2\pi r_{+}^{2}}\left[M\left(r_{+}\right)-\frac{Q^{2}\left(r_{+}\right)}{r_{+}}-r_{+}\left.\frac{dM}{dr}\right|_{r=r_{+}}+Q\left(r_{+}\right)\left.\frac{dQ}{dr}\right|_{r=r_{+}}\right].\label{eq:temp1}
\end{eqnarray}

The entropy in terms of the area of the horizon is given by the well
known relation

\begin{equation}
S=\frac{A}{4}=\pi r_{+}^{2}
\end{equation}
 and therefore, the Komar energy (\ref{eq:KomarEnergy}) at the event
horizon becomes

\begin{equation}
E=2\pi r_{+}^{2}T=2ST.\label{komar_ener}
\end{equation}

Using the value $r_{\pm}=M_{0}\pm\sqrt{M_{0}^{2}-Q_{0}^{2}}$ as a
first approximation of the horizons (\ref{eq:horizonsDef}) and putting
them into the incomplete gamma functions of relations (\ref{eq:Q})
and (\ref{eq:M}) one obtains 

\noindent 
\begin{equation}
r_{\pm}=M_{\pm}\pm\sqrt{M_{\pm}^{2}-Q_{\pm}^{2}}\label{eq:horizons}
\end{equation}
where we have defined

\begin{equation}
M_{\pm}=M_{0}\left[\varepsilon\left(\frac{M_{0}\pm\sqrt{M_{0}^{2}-Q_{0}^{2}}}{2\ell}\right)-\frac{M_{0}\pm\sqrt{M_{0}^{2}-Q_{0}^{2}}}{\sqrt{\pi}\ell}\exp\left(-\frac{\left(M_{0}\pm\sqrt{M_{0}^{2}-Q_{0}^{2}}\right)^{2}}{4\ell^{2}}\right)\right]
\end{equation}

\begin{equation}
Q_{\pm}=Q_{0}\sqrt{\varepsilon^{2}\left(\frac{M_{0}\pm\sqrt{M_{0}^{2}-Q_{0}^{2}}}{2\ell}\right)-\frac{\left(M_{0}\pm\sqrt{M_{0}^{2}-Q_{0}^{2}}\right)^{2}}{\sqrt{2\pi}\ell^{2}}\exp\left(-\frac{\left(M_{0}\pm\sqrt{M_{0}^{2}-Q_{0}^{2}}\right)^{2}}{4\ell^{2}}\right)}
\end{equation}
and $\varepsilon\left(x\right)$ is the Gauss error function,

\begin{equation}
\varepsilon\left(x\right)=\frac{2}{\sqrt{\pi}}\int_{0}^{x}e^{-u^{2}}du.
\end{equation}

For a large value of its argument (i.e. large masses), function $\varepsilon$
tends to unity while the exponential term goes to zero, giving the
classical Reissner-Nordström horizons $r_{\pm}\rightarrow r_{RN\pm}=M_{0}\pm\sqrt{M_{0}^{2}-Q_{0}^{2}}$.

Using the same value as a first approximation for the event horizon
in the Hawking temperature (\ref{Hawk_temp}) one obtains \cite{mehdipour10}

\begin{equation}
T\approx\frac{1}{4\pi}\frac{r_{+}-r_{-}}{r_{+}^{2}}.\label{eq:temp2}
\end{equation}

\noindent This approximation permit us to write the Komar energy at
the horizon, using Eqs.(\ref{komar_ener}), (\ref{eq:temp2}) and
(\ref{eq:horizons}), as

\noindent 
\begin{equation}
E=2\pi r_{+}^{2}T=\frac{r_{+}-r_{-}}{2}
\end{equation}
\begin{eqnarray}
E & = & \frac{1}{2}\left[M_{+}+M_{-}+\sqrt{M_{+}^{2}-Q_{+}^{2}}-\sqrt{M_{-}^{2}-Q_{-}^{2}}\right].\label{komar1c}
\end{eqnarray}

By considering the behavior of the functions $M_{\pm}$ and $Q_{\pm}$,
it is easy to see that the limit of large masses of (\ref{komar1c}),
as well as taking the limit $\ell\rightarrow0$, recover the Reissner-Nordström
energy and for $Q_{0}=0$ it gives the result of Banerjee and Gangopadhyay
\cite{banerjee10} for the noncommutative Schwarzschild black hole
with the usual $E=M_{0}$ that let us identify the quantity $M_{0}$
as the total mass of the black hole and $Q_{0}$ as its total electric
charge. 

With a similar procedure, the entropy can be approximated by 
\begin{eqnarray}
S & = & \pi r_{+}^{2}\approx\pi\left(M_{+}+\sqrt{M_{+}^{2}-Q_{+}^{2}}\right)^{2},\label{eq:entropy1}
\end{eqnarray}
which give in the limit of large masses, or in the limit $\ell\rightarrow0$,
the usual result for the Reissner-Nordström black hole, $S\rightarrow S_{RN}=\pi\left(M_{0}+\sqrt{M_{0}^{2}-Q_{0}^{2}}\right)^{2}$.

Using Eqs. (\ref{eq:Q}) and (\ref{eq:M}) and the property of the
gamma function

\begin{equation}
\frac{\partial}{\partial u}\gamma\left(\frac{a}{b},u\right)=e^{-u}u^{-1+\frac{a}{b}}
\end{equation}
to perform the derivatives, the Komar energy (\ref{eq:KomarEnergy})
for this spacetime yields 
\begin{eqnarray}
E & = & M(r)-\frac{Q^{2}\left(r\right)}{r}-\frac{M_{0}}{2\sqrt{\pi}}\frac{r^{3}}{\ell^{3}}e^{-\frac{r^{2}}{4\ell^{2}}}\nonumber \\
 &  & +\frac{Q_{0}^{2}}{2\pi}\left[\frac{2}{\ell}e^{-\frac{r^{2}}{4\ell^{2}}}\gamma\left(\frac{1}{2},\frac{r^{2}}{4\ell^{2}}\right)-\frac{1}{\sqrt{2}\ell}\gamma\left(\frac{1}{2},\frac{r^{2}}{2\ell^{2}}\right)+\frac{\sqrt{2}}{\ell}\gamma\left(\frac{3}{2},\frac{r^{2}}{4\ell^{2}}\right)-\frac{r}{\ell^{2}}e^{-\frac{r^{2}}{2\ell^{2}}}+\frac{\sqrt{2}}{4}\frac{r^{3}}{\ell^{4}}e^{-\frac{r^{2}}{4\ell^{2}}}\right].\label{komar_comp}
\end{eqnarray}

Using the long distance approximations for the gamma functions

\begin{equation}
\gamma\left(\frac{3}{2},\frac{r^{2}}{4\ell^{2}}\right)\simeq\frac{\sqrt{\pi}}{2}-\frac{r}{2\ell}e^{-r^{2}/4\ell^{2}}
\end{equation}

\begin{equation}
\gamma\left(\frac{1}{2},\frac{r^{2}}{2\ell^{2}}\right)\simeq\sqrt{\pi}-\sqrt{2}\ell\frac{e^{-r^{2}/2\ell^{2}}}{r}
\end{equation}

\begin{equation}
\gamma\left(\frac{1}{2},\frac{r^{2}}{4\ell^{2}}\right)\simeq\sqrt{\pi}-2\ell\frac{e^{-r^{2}/4\ell^{2}}}{r}
\end{equation}

we obtain the relation

\begin{align}
M_{0}-\frac{Q_{0}^{2}}{r}= & 2TS+\frac{M_{0}}{\sqrt{\pi}}\frac{r}{\ell}e^{-\frac{r^{2}}{4\ell^{2}}}\left(1+\frac{r^{2}}{2\ell^{2}}\right)\nonumber \\
 & +\frac{Q_{0}^{2}}{\pi r}\left[e^{-\frac{r^{2}}{2\ell^{2}}}\left(\frac{5}{2}+\frac{r^{2}}{2\ell^{2}}+\frac{4\ell^{2}}{r^{2}}\right)-e^{-\frac{r^{2}}{4\ell^{2}}}\left(4\sqrt{\pi}\frac{\ell}{r}+\sqrt{\pi}\frac{r}{\ell}+\frac{\sqrt{2}}{4}\frac{r^{2}}{\ell^{2}}+\frac{\sqrt{2}}{8}\frac{r^{4}}{\ell^{4}}\right)\right].\label{eq:Smarr}
\end{align}

Since $M_{0}$ and $Q_{0}$ have been identified as the mass and charge
of the black hole, Eq.(\ref{eq:Smarr}) corresponds to the generalization
of the \textit{\emph{Smarr formula for the noncommutative charged
black hole}}. Note that this relation deviates from the usual one
(\ref{eq:generalSmarr}) by the two last terms in the right hand side,
but it is clear that in the limit $\ell\rightarrow0$ these terms
disappear. In the case $Q_{0}=0$ we recover the relation for the
noncommutative Schwarzschild black hole presented in \cite{banerjee10,samanta,majhimodak}.

\section{\noindent Conclusion}

We have computed the Komar energy for a charged black hole inspired
in noncommutative geometry and its asymptotic limit that let us identify
the constant $M_{0}$ as its total mass and $Q_{0}$ as its electric
charge. With these results, we obtain the noncommutative version of
the Smarr formula (\ref{eq:Smarr}) which show a deformation from
the usual relation and the new terms depend on the noncommutative
parameter $\ell$. 

\emph{Acknowledgements}

This work was supported by the Universidad Nacional de Colombia. Hermes
Project Code 13038.


\begin{thebibliography}{References}
\bibitem{Beken3} J.D.Bekenstein, Phys. Rev. D \textbf{7}, 2333 (1973) 

\bibitem{Hawking1} S.W.Hawking, Nature \textbf{248}, 30 (1974)

\bibitem{Hawking2} S.W.Hawking, Commun. Math. Phys. \textbf{43},
199 (1975)

\bibitem{paddy}D. Kothawala, T. Padmanabhan, S. Sarkar, Phys. Rev.
D \textbf{78}, 104018 (2008)

\bibitem{padma}T. Padmanabhan, Class. Quant. Grav. \textbf{21}, 4485
(2004)

\bibitem{komar}A. Komar, Phys. Rev. \textbf{113}, 934 (1959). 

\bibitem{wald}R.M. Wald, ``General Relativity\textquotedbl{}, Chicago,
U.S.A : University Press (1984)

\bibitem{smarr} L. Smarr, Phys. Rev. Lett. \textbf{30}, 71 (1973),
{[}Erratum-ibid. 30, 521 (1973){]}. 

\bibitem{rbstat}R. Banerjee, B.R. Majhi, Phys. Rev. D\textbf{81},
124006 (2010) 

\bibitem{bmms}R. Banerjee, B.R. Majhi, S.K. Modak, S. Samanta, Phys.
Rev. D\textbf{82}, 124002 (2010) 

\bibitem{ms}S.K. Modak, S. Samanta. arXiv:1006.3445 {[}gr-qc{]}

\bibitem{Poisson}Eric Poisson, \textquotedblleft{}A Relativist\textquoteright{}s
Toolkit; The Mathematics of Black Hole Mechanics\textquotedblright{}
Cambridge University Press (2004).

\bibitem{NC1}A. Smailagic and E. Spallucci, Phys. Rev. D \textbf{65},
107701 (2002)

\bibitem{NC2}A. Smailagic and E. Spallucci, J. Phys. A \textbf{35},
L363 (2002)

\bibitem{NC3}A. Smailagic and E. Spallucci, J. Phys. A \textbf{36},
L467 (2003)

\bibitem{NC4}A. Smailagic and E. Spallucci, J. Phys. A \textbf{36},
L517 (2003)

\bibitem{NC5}A. Smailagic and E. Spallucci, J. Phys. A \textbf{37},
7169 (2004)

\bibitem{spal} P.Nicolini, A.Smailagic, E.Spallucci, Phys. Lett.
B \textbf{632}, 547 (2006) 

\bibitem{spalreview} P.Nicolini, Int. J. Mod. Phys A \textbf{24},
7, 1229 (2009)

\bibitem{spal2}S. Ansoldi, P. Nicolini, A. Smailagic, E. Spallucci,
Phys. Lett. B \textbf{645}, 261 (2007)

\bibitem{sunfgs}S. Gangopadhyay, F.G. Scholtz, Phys. Rev. Lett. \textbf{102},
241602 (2009)

\bibitem{sunPLB}R. Banerjee, S. Gangopadhyay, S.K. Modak, Phys. Lett.
B \textbf{686}, 181 (2010)

\bibitem{rbreview} R. Banerjee, B. Chakraborty, S. Ghosh, P. Mukherjee,
S. Samanta, Found. Phys. 39: 1297, (2009)

\bibitem{moyal1}H. Weyl, Z. Phys. \textbf{46}, 1 (1927)

\bibitem{moyal2}E. Wigner, Phys. Rev. \textbf{40}, 749 (1932)

\bibitem{moyal3}J. E. Moyal, Proc. Camb. Phil. Soc. \textbf{45},
99 (1949)

\bibitem{banerjee10}R. Banerjee and S. Gangopadhyay. Gen. Rel. Grav.
\textbf{43}, 3201-3212 (2011) 

\bibitem{mehdipour10}S. H. Mehdipour. Int. J. Mod. Phys. A \textbf{25},
5543-5555 (2010) 

\bibitem{samanta}R. Banerjee, B.R. Majhi, S. Samanta, Phys. Rev.
D \textbf{77}, 124035 (2008) 

\bibitem{majhimodak}R. Banerjee, B.R. Majhi, S.K. Modak, Class. Quant.
Grav., \textbf{26}, 085010 (2009)

\bibitem{pedro}E. Elizalde, J.S. Pedro, Phys. Rev. D \textbf{78},
061501 (2008)

\bibitem{hogan}C.J. Hogan, Phys. Rev. D \textbf{77}, 104031 (2008)

\bibitem{chaichian} M. Chaichian, A. Tureanu, G. Zet, Phys. Lett.
B 660: 573, (2008); {[}arXiv:0710.2075 {[}hep-th{]}{]}. 

\bibitem{anirban}P. Mukherjee, A. Saha, Phys. Rev. D 77: 064014,
(2008); {[}arXiv:0710.5847 {[}hep-th{]}{]}. \end{thebibliography}
\end{document}